\documentclass[smallextended]{svjour3}       

\smartqed  
\usepackage{mathptmx}      

\usepackage{wrapfig}
\usepackage{amssymb}
\usepackage{graphicx}
\usepackage{subfigure}
\usepackage{tikz}
\usetikzlibrary{arrows}
\usetikzlibrary{shapes}
\usetikzlibrary{decorations.markings}
\usepackage{algorithm}
\usepackage{algorithmic}
\usepackage{amsmath}			
\usepackage{amssymb}			
\usepackage{float}
\usepackage{url}
\usepackage{hyperref}
\hypersetup{colorlinks=true}
\hypersetup{breaklinks=true}

\long\def\comment#1{}

\tikzstyle{edge} = [draw,line width=1pt,->,>=stealth',shorten >=1pt,black]

\journalname{Algorithmica}

\begin{document}
	
	\title{Complexity issues in some  clustering problems in combinatorial circuits \thanks{V. Mkrtchyan and K. Subramani are supported, in part, by the Air Force of Scientific Research through Award FA9550-12-1-0199.} \thanks{K. Subramani is supported by the National Science Foundation through Award CCF-1305054.}
	}
	
	\subtitle{When logic replication is not allowed}
	
	\titlerunning{On clustering without replication}        
	
	\author{Z. Donovan \and V. Mkrtchyan \and K. Subramani}
	
	\authorrunning{Z. Donovan \and V. Mkrtchyan \and K. Subramani} 
	
	\institute{Zola Donovan \at
		Department of Mathematics, West Virginia University, Morgantown, WV, USA  \\
		\email{zdonovan@mix.wvu.edu}           
		\and
		Vahan Mkrtchyan \at
		LDCSEE, West Virginia University, Morgantown, WV, USA \\
		\email{vahanmkrtchyan2002@ysu.am}
		\and
		K. Subramani \at
		LDCSEE, West Virginia University, Morgantown, WV, USA \\
		\email{K.Subramani@mail.wvu.edu}
	}
	
	\date{Received: date / Accepted: date}

	\maketitle

	\begin{abstract}
		The modern integrated circuit is one of the most complex products that has been engineered to-date. It continues to grow in complexity as the years progress. As a result, very large-scale integrated (VLSI) circuit design now involves massive design teams employing state-of-the art computer-aided design (CAD) tools. One of the oldest, yet most important CAD problems for VLSI circuits is physical design automation, where one needs to compute the best physical layout of millions to billions of circuit components on a tiny silicon surface \cite{Lim08}. The process of mapping an electronic design to a chip involves a number of physical design stages, one of which is clustering. In this paper, we focus on problems in clustering which are critical for more sustainable chips. The clustering problem in combinatorial circuits alone is a source of multiple models. In particular, we consider the problem of clustering combinatorial circuits for delay minimization, when logic replication is not allowed ({\sc CN}). The problem of delay minimization when logic replication is allowed ({\sc CA}) has been well studied, and is known to be solvable in polynomial-time \cite{Wong1}. However, unbounded logic replication can be quite expensive. Thus, {\sc CN} is an important problem. We show that selected variants of {\sc CN} are {\bf NP-hard}. We also obtain approximability and inapproximability results for these problems. A preliminary version of this paper appeared in \cite{Don15}.
		
	\end{abstract}

	\section{Introduction}
	\label{sec:intro}
	
	In this paper, we consider the problem of clustering combinatorial circuits for delay minimization when logic replication is not allowed ({\sc CN}). Combinatorial circuits implement Boolean functions, and produce a unique output for every combination of input signals \cite{Kos04}. The gates and their interconnections in the circuit represent implementations of one or more Boolean function(s). The Boolean functions are realized by the assignment of the gates to chips.
	
	Due to manufacturing process and capacity constraints, it is generally not possible to place all of the circuit elements in one chip. Consequently, the circuit must be partitioned into clusters, where each cluster represents a chip in the overall circuit design. The circuit elements are assigned to clusters, while satisfying certain design constraints (e.g., area capacity) \cite{Wong1}.
	
	Gates and their interconnections usually have delays. The delays of the interconnections are determined by the way the circuit is clustered. Intra-cluster delays are associated with the interconnections between gates in the same cluster. Inter-cluster delays are associated with the interconnections between gates in different clusters. The delay along a path from an input to an output is the sum of the delays of the gates and interconnections on the respective paths. The delay of the overall circuit, with respect to its clustering, is the maximum delay among all paths that connect an input to any output in the circuit.
	
	\begin{wrapfigure}{R}{0.4\textwidth}
		\begin{center}
			\begin{tikzpicture}[->,>=stealth',shorten >=1pt,auto,node distance=1.35cm, thick,
			main node/.style={circle,fill=blue!20,draw,font=\sffamily\Large\bfseries},
			blank node/.style={circle,fill=white!0,draw=none,font=\sffamily\Large\bfseries,opacity=0}]
			
			\node[main node] (a) [label=left:$a$] {};
			\node[main node] (b) [label=left:$b$] [below of =a] {};
			\node[main node] (c) [label=left:$c$] [below right of =b] {};
			\node[blank node] (i) [above right of =c] {};
			\node[main node] (g) [label=right:$g$] [above right of =i] {};
			\node[main node] (f) [label=right:$f$] [below right of =g] {};
			\node[main node] (e) [label=right:$e$] [below of =f] {};
			
			\path[every node]
			(a)	edge  (c)
			edge  (g)
			edge  (e)

			(b)	edge (g)
			edge (e)
			edge (f)
			
			(c)	edge (g)
			edge (e)
			
			(g)	edge (f)
			edge (e)
			;
			
			\end{tikzpicture}
			
		\end{center}
		
		\caption{A DAG representing a combinatorial network with two sources and two sinks.} \label{DAGexample}
	\end{wrapfigure}
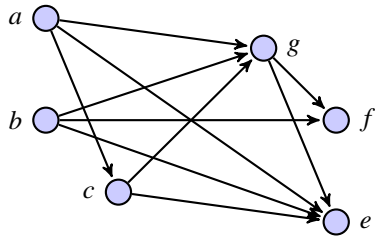

	The problem of clustering combinatorial circuits for delay minimization, when logic replication is allowed ({\sc CA}), is well studied. It arises frequently in VLSI design. In {\sc CA}, the goal is to find a clustering of a circuit for which the delay of the overall circuit is minimized. {\sc CA} has been shown to be solvable in polynomial-time \cite{Wong1}. However, unbounded replication can be quite expensive. As systems become increasingly more complex, the need for clustering without logic replication is crucial. It follows that {\sc CN} is an important problem in VLSI design.
	
	In this paper, we consider several variants of {\sc CN}. We prove {\bf NP-hardness} results for these variants. We design an approximation algorithm for one of them. We also obtain inapproximability results.
	
	The rest of this paper is organized as follows: \comment{We give the necessary graph-theoretic definitions in Section \ref{sec:prelim}.} The problem is formally described in Section \ref{sec:sop}.
	We then examine related work in Section \ref{sec:relwork}. In Section \ref{sec:comp}, we give some hardness results for the clustering problem. We also show that our hardness results imply inapproximability.
	In Section \ref{sec:ApproxResults}, we propose an approximation algorithm for solving the clustering problem, when the gates are unweighted and the cluster capacity, $M$, is $2$.
	We conclude the paper with Section \ref{sec:conc} by summarizing our main results, and identifying avenues for future work.
	
	\section{Statement of Problems}
	\label{sec:sop}
	
	In this section, we formally describe the problem studied in this paper. We start with graph preliminaries. Next, we formulate the problem using the language of combinatorial circuits. Finally, we represent such circuits as directed acyclic graphs and formulate the main problem using graph-theoretic terminology.
	
	\begin{subsection}{Graph Preliminaries}
		\label{sec:prelim}
		
		In this subsection, we define the main graph-theoretic concepts that are used in the paper.
		
		Graphs considered in this paper do not contain loops or parallel edges. The degree of a vertex $v$ of a graph $G$ is the number of edges of $G$ incident with $v$. 
		
		A path of $G$ is a sequence $P=v_0, e_1, v_1,..., e_l, v_l$, where $v_0, v_1, ..., v_l$ are vertices of $G$, $e_1, ..., e_l$ are edges of $G$, and $e_j=(v_{j-1}, v_j)$, $1\leq j \leq l$. $l$ is called the  length of the path $P$, and sometimes we say that $P$ is an $l$-path of $G$. The edge $e_{\lceil \frac{l}{2} \rceil}$ is called a central edge of $P$. $G$ is connected if any two vertices of $G$ are joined by a path of $G$. $P$ is said to be a cycle, if $v_0=v_l$.
		
		A directed path of a directed graph $D$ is a sequence $Q=v_0, e_1, v_1,..., e_l, v_l$, where $v_0, v_1, ..., v_l$ are vertices of $D$, $e_1, ..., e_l$ are edges of $D$, and $e_j=(v_{j-1}, v_j)$, $1\leq j \leq l$. $l$ is called the length of the path $Q$, and sometimes we say that $Q$ is a directed $l$-path of $D$. If $v_0=v_l$, then $Q$ is called a directed cycle. $D$ is said to be a directed acyclic graph (DAG), if it contains no directed cycles.
		
		A cluster is defined as a subset of the vertices of a graph. If $C$ is a cluster in a graph, then an edge is said to be a cut-edge if it connects a vertex of $C$ to a vertex from $V\backslash C$. The degree of $C$ is the number of cut-edges incident with a vertex in $C$.
		
		The fanin and fanout of a vertex are the number of arcs 
        that enter and leave the vertex, respectively. A source represents a vertex with fanin equal to $0$, and a sink represents a vertex with fanout equal to $0$. As the example from Figure \ref{DAGexample} shows, a DAG may have more than one source and more than one sink.
		
		Let $\mathcal{I}$ and $\mathcal{O}$ be the set of sources and sinks of $G$, respectively. Notice that $\mathcal{I}=\{a, b\}$ and $\mathcal{O}=\{e,f\}$ in the DAG in Figure \ref{DAGexample}; $C_1=\{a, c, g\}$ and  \mbox{$C_2=\{b, e, f\}$} represent a pair of disjoint clusters.
	\end{subsection}
	
	\begin{subsection}{Formulation of the problem using the language of combinatorial circuits}
		\label{sec:combcirform}
		In general, each gate in a circuit has an associated delay \cite{Murg1}. In the model that we consider in this paper, each interconnection has one of the following types of delays: (1) an intra-cluster delay, $d$, when there is an interconnection between two gates in the same cluster, or (2) an inter-cluster delay, $D$, when there is an interconnection between two gates in different clusters.
		
		Note that $D>>d$, so inter-cluster delays typically dominate in all delay calculations.
		
		The delay along a path from an input to an output is the sum of the delays of the gates and interconnections that lie on the path. The delay of the overall circuit is the maximum delay among all source to sink paths in the circuit.
		
		Technology and design paradigms impose a number of constraints on the clustering of a circuit. So, a clustering is feasible if all clusters obey the imposed constraints. Constraints can be either monotone or non-monotone. In \cite{Lawvlsi1}, the following definition is given:
		
		\begin{definition} A constraint is said to be monotone if and only if any connected subset of gates in a feasible cluster is also feasible.
		\end{definition}
		
		A typical constraint includes capacity (a monotone constraint), which is a fixed constant $M$, denoting an upper-bound on the number of gates allowed in a cluster
        .
		
		In {\sc CN}, a clustering partitions the circuits into disjoint subsets.
		
		A clustering algorithm tries to achieve one or both of the following goals, subject to one or more constraints:
		\begin{itemize}
			\item [(1)] The delay minimization through the circuit \cite{Wong1}\comment{, \cite{Murg1}, \cite{Cong1}, \cite{Lawvlsi1}}.
			\item [(2)] The minimization of the total number of cut-edges \cite{HG}\comment{, \cite{LKC}, \cite{SK}, \cite{DD}, \cite{MW}, \cite{BHB}}.
		\end{itemize}
		
		In this paper, we study {\sc CN} under the delay model described as follows:
		\begin{enumerate}
			\item Associated with every gate $v$ of the circuit, there is a delay $\delta(v)$ and a size $w(v)$.
			\item The delay of an interconnection between two gates within a single cluster is $d$.
			\item The delay of an interconnection between two gates in different clusters is $D$, where $D>>d$.
		\end{enumerate}
		
		The size of a cluster is the sum of the sizes of the gates in the cluster. 
		
		The precise formulation of the problem is as follows: \\
		
		{\sc CN}: Given a combinatorial circuit, with each gate having a size and a delay, intra- and inter-cluster delays $d$ and $D$, respectively, and a positive integer $M$ called cluster capacity, the goal is to partition the circuit into clusters such that
		\begin{enumerate}
			\item  The size of each cluster is bounded by $M$,
			\item The delay of the circuit is minimized.
		\end{enumerate}
		\bigskip
		
		A combinatorial circuit can be represented as a directed graph $G=(V,E)$, with vertex-set $V$ and edge-set $E$, such that $G$ has no directed cycles. In $G$, each vertex $v \in V$ represents a gate, and each edge $(u,v) \in E$ represents an interconnection between gates $u$ and $v$.
		
		Given a clustering of the combinatorial circuit, the delays on the interconnections between gates  induce an edge-length function $l: E(G)\rightarrow \{d, D\}$ of $G$. The weight of a cluster is the sum of the weights of the vertices in the cluster.
	\end{subsection}
	
	\begin{subsection}{Formulation of the problem using graph-theoretic terms}
		\label{sec:gtform}
		
		In the rest of the paper, we focus on a graph-theoretic formulation of {\sc CN}. We employ the following notations and concepts: The length of a path $P$ in $G$ is calculated as the sum of all delays of vertices and edge-lengths of edges of $P$. $X$ below can be either $W$, which means that the vertices are weighted, or $N$, which means that the vertices are unweighted. $M$ is the cluster capacity. $\Delta$ is the maximum \comment{degree of the underlying simple graph} number of arcs entering or leaving any vertex of the DAG.
		
		{\sc CN} is formulated (graph-theoretically) as follows:\\
		
		{\sc CN}$\langle X, M, \Delta \rangle$: Given a DAG $G=(V,E)$, with vertex-weight function $w: V \rightarrow \mathbb{N}$, delay function $\delta: V \rightarrow \mathbb{N}$, constants $d$ and $D$, and a cluster capacity $M$, the goal is to partition $V$ into clusters such that
		\begin{enumerate}
			\item  The weight of each cluster is bounded by $M$,
			\item The maximum length of any path from a source to a sink of $G$ is minimized.
		\end{enumerate}
		
		A clustering of $G$, such that the weight of each cluster is bounded by $M$, is called feasible. Given a feasible clustering of $G$, one can consider the corresponding edge-length function $l: E(G)\rightarrow \{d, D\}$ of $G$. A maximum length path (with respect to $l$) from a source to a sink of $G$ is called an optimal path. A clustering of $G$ is optimal, if the length of an optimal path is the smallest. An optimal path with respect to an optimal clustering is called a critical path.
		
		\begin{figure}[ht]
			\begin{center}
				\begin{tikzpicture}[->,>=stealth',shorten >=1pt,auto,node distance=1.25cm, thick,
				main node/.style={circle,fill=blue!20,draw,font=\sffamily\small\bfseries, scale=.8}]
				
				\node[main node] (s) {$s$};
				\node[main node] (a) [above right of =s] {$a$};
				\node[main node] (b) [below right of =s] {$b$};
				\node[main node] (c) [right of =a] {$c$};
				\node[main node] (e) [right of =b] {$e$};
				\node[main node] (t) [below right of =c] {$t$};

				\path[every node]
				(s)	edge  (a)
				edge  (b)
				
				(a)	edge (c)
				edge (e)
				
				(b)	edge (c)
				edge (e)
				
				(c)	edge (t)
				
				(e)	edge (t)
				
				;
				
				\end{tikzpicture} \hspace{20pt}
				\begin{tikzpicture}[->,>=stealth',shorten >=1pt,auto,node distance=1.25cm, thick,
				main node/.style={circle,fill=blue!20,draw,font=\sffamily\small\bfseries, scale=.8}]
				
				\node[main node] (s) {$s$};
				\node[main node] (a) [above right of =s] {$a$};
				\node[main node] (b) [below right of =s] {$b$};
				\node[main node] (c) [right of =a] {$c$};
				\node[main node] (e) [right of =b] {$e$};
				\node[main node] (t) [below right of =c] {$t$};

				\path[every node]
				(s)	edge  (a)
				edge  (b)
				
				(a)	edge (c)
				edge (e)
				
				(b)	edge (c)
				edge (e)
				
				(c)	edge (t)
				
				(e)	edge (t)
				
				;
				
				\draw[rotate=45] (.5, 0) ellipse (1.125 cm and .25 cm);
				\draw[rotate=0] (1.25, -.7) ellipse (1.125 cm and 0.25cm);
				\draw[rotate=-45] (1.25, 1.7) ellipse (1.125 cm and 0.25cm);
				
				\end{tikzpicture}
				
			\end{center}
			
			\caption{An example of a DAG and its clustering. \label{Networkexample2}}
		\end{figure}
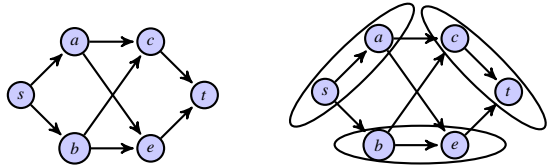
		
		\begin{wrapfigure}[14]{r}{7cm}
			\vspace{-.4in}
			\tikzstyle{edge} = [draw,line width=1pt,->,>=stealth',shorten >=1pt,black]
			\centering
			\begin{tikzpicture}[-,>=stealth',shorten >=.5pt,auto,node distance=1cm, thin,
			boxed node/.style={rectangle,draw,font=\sffamily\small\bfseries, scale=.5},
			blank node/.style={font=\sffamily\LARGE\bfseries, scale=.5}]
			
			\node[blank node] (0) at (0,3) {$CN\langle W, M, \Delta \rangle$};
			\node[blank node] (1a) at (-3,1) {$CN\langle N, M, \Delta \rangle$};
			\node[blank node] (1b) at (0,1) {$CN\langle W, 2, \Delta \rangle$};
			\node[blank node] (1c) at (3,1) {$CN\langle W, M, 3 \rangle$};
			\node[blank node] (2a) at (-3,-1) {$CN\langle N, 2, \Delta \rangle$};
			\node[blank node] (2b) at (0,-1) {$CN\langle N, M, 3 \rangle$};
			\node[blank node] (2c) at (3,-1) {$CN\langle W, 2, 3 \rangle$};
			\node[blank node] (3) at (0,-3) {$CN\langle N, 2, 3 \rangle$};

			\foreach \source/ \dest in {0/1a, 0/1b, 0/1c, 1a/2a, 1a/2b, 1b/2a, 1b/2c, 1c/2b, 1c/2c, 2a/3, 2b/3, 2c/3}
			\path[edge] (\source) -- (\dest);
			;
			
			\end{tikzpicture}\caption{Cases of the delay minimization problem that we plan to investigate.} \label{fig:lattice}
		\end{wrapfigure}
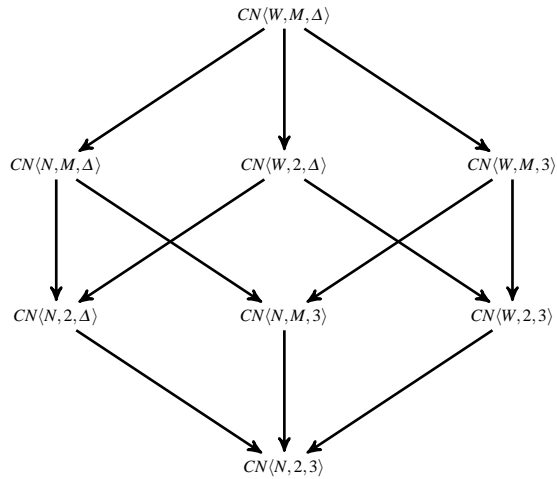
		
		In Figure \ref{Networkexample2}, we consider a simple example of a clustering of a combinatorial circuit represented by a DAG, where logic replication is not allowed.
		In this example, the weights and delays of all vertices are equal to $1$ (i.e., $\delta(v)=1$ and $w(v)=1$ for all vertices $v$ in the DAG); the upper bound for the weight of the cluster is $M=2$; the intra-cluster delay is $d=1$; and, the inter-cluster delay is $D=2$. It can be easily seen that the partition $\Sigma=\{\{s,a\}, \{b,e\}, \{c,t\} \}$ forms a feasible clustering such that the length of the optimal path is $9$. Moreover, it can be checked that this clustering is optimal.

		We investigate the delay minimization problem for the cases shown in Figure \ref{fig:lattice}. In particular, our goal is to obtain reductions among these problems.
		
		In this paper, we focus on a restriction of {\sc CN}$\langle X, M, \Delta \rangle$ , when $\delta(v)=0$ for any vertex $v$ of $G$.
		
	\end{subsection}
	
	The main contributions of this paper are as follows:
	\begin{enumerate}
		\item Establishing the {\bf NP-hardness} of {\sc CN}$\langle W, M, \Delta \rangle$ and several of its variants (Section \ref{sec:comp}).
		\item Design and analysis of a $3$-approximation algorithm for {\sc CN}$\langle N, 2, \Delta \rangle$ (Section \ref{sec:ApproxResults}).
		\item Proof of inapproximability for several variants of {\sc CN}$\langle W, M, \Delta \rangle$. \comment{Namely, that the problem does not admit a $(2-\varepsilon)$-approximation algorithm for every $\varepsilon >0$, unless {\bf P=NP}} (Section \ref{sec:comp}).
	\end{enumerate}

	\section{Related Work}
	\label{sec:relwork}
	
	In this section, we describe some related work in the literature.
	
	\comment{There has been considerable attention given to optimizing VLSI circuits for power for more than a decade~\cite{Devadas:1995:SOT:217474.217536},~\cite{573196}.  Life-cycle studies of computing systems~\cite{Yao2010},~\cite{doi:10.1021/es303012r},~\cite{Teehan13},~\cite{AppleEnv15}, including previous work of the PIs~\cite{6691120},~\cite{6604497},~\cite{Jones15}, have pinpointed manufacturing impacts of ICs as a problem of growing importance.  To our knowledge there are no CAD techniques that attempt to reduce overall environmental impact of designs that include manufacturing impacts.}
	
	In \cite{Lawvlsi1}, the authors present an exact polynomial-time algorithm for {\sc CA}. The problem is solved under the so-called unit delay model \cite{Lawvlsi1}.
	
	A more general delay model is presented in \cite{Murg1}. The problem of disjoint clustering for minimum delay under the area or pin constraint is shown to be intractable in \cite{Murg1}. To minimize the delay, the authors propose an algorithm which constructs a clustering. This algorithm achieves the optimal delay under specific conditions.
	
	In \cite{Wong1}, {\sc CA} is considered under the more general delay model proposed in \cite{Murg1}. However, \cite{Wong1} presents a different polynomial-time algorithm. Their heuristic is shown to always find an optimal clustering under any monotone clustering constraint.
	
	Similar to \cite{Murg1}, the problem of disjoint clustering for minimum delay under the area or pin constraint is also shown in \cite{Kag03} to be intractable. However, an improved heuristic is proposed in \cite{Kag03}. The authors also share comparative experimental results which show that a decrease in clusters generally leads to an increase in maximum delay.
	
	In \cite{MW}, the authors propose an efficient network-flow based algorithm which determines an optimal partitioning of the circuit. Using the least amount of replication, the optimal partitioning separates the nodes of the circuit into two subsets with the smallest cut size. The algorithm presented in \cite{MW} is also applicable to size-constrained partitioning.
	
	\cite{Shan11} and \cite{Shan14} explore the advantage of evolutionary algorithms aimed at reducing the delay and area in partitioning and floorplanning. In turn, this would reduce the wirelength. A hybrid of the evolutionary algorithms are used to find optimal solutions to VLSI physical design problems.
	
	In \cite{Moi15}, the authors present an algorithm for simultaneous multilayer interconnect spacing. While satisfying maximum delay constraints, their unique algorithm guarantees to minimize the total dynamic power dissipation caused by an interconnect.
	
	In \cite{Su10}, adjustable delay buffers (ADBs) are used to minimize clock skew under different power modes. The ADBs have delays which can be tuned or adjusted. When the positions of some fixed number of ADBs are assumed to be predetermined, the authors propose a linear-time optimal algorithm. This algorithm assigns the values of the ADBs so as to minimize clock skew among all possible ADB assignments. In this case, there is a possibility of latency penalty. They also propose a modified algorithm to find an optimal solution with no latency penalty. Additionally, they give an efficient heuristic for finding good ADB positions.
	
	Similar to \cite{Su10}, the author of \cite{Kao15} studies the use of ADBs to minimize clock skew under different power modes. In order to generate zero clock skew in a given clock tree, they start by assigning ADB positions. If the number of ADBs assigned do not meet the constraints of the previous solution, they use a bottom-up approach for removing ADBs to minimize clock skew while satisfying all constraints.
	
	\cite{Sub10} examines the methods used to solve bi-criterion VLSI circuit partitioning problems. The authors present a hybrid genetic algorithm (GA) which employs the Taguchi method for local search. They test their hybrid algorithm with a variety of benchmarks circuits, and found it superior in comparison to the standard GA and tabu search algorithms reported in the literature.
	
	A routability-driven clustering technique for area and power reduction in clustered FPGAs is presented in \cite{SinghM02}. This technique uses a cell connectivity metric to identify seeds for efficient clustering. Effective seed selection, coupled with an interconnect-resource aware clustering and placement, can have a remarkable impact on circuit routability. It leads to better device utilization, reduction in power consumption and savings in area  \cite{SinghM02}. Additionally, routing area is reduced by $35\%$.
	The authors also show that their clustering technique can reduce the overall device power usage by an average of $13\%$.
	
	In \cite{Manik1}, effective circuit partitioning techniques are employed by using
	clustering algorithms. The technique presented in \cite{Manik1} uses the circuit netlist in order to cluster
	the circuit in partitioning steps. It also minimizes the interconnection distance
	with the required iteration level. For the standard benchmark circuits the well-known clustering algorithms like $K$-Mean, $Y$-Mean,
	$K$-Medoid are performed. The results obtained in \cite{Manik1}
	show that the proposed techniques improve the delay.  They also minimize the area
	by reducing the interconnection distance.
	
	The multiway partition problem \comment{discussed in Section \ref{clus} }remains {\bf NP-hard}, even when the input hypergraph is an unweighted graph, and there is no  restriction on the sizes of clusters. If the number of clusters  is fixed (say $r$), then there is an algorithm that runs in time $O(n^{r^{2}})$ that solves this restriction exactly \cite{Goldschmidt1}. Here $n$ is the number of vertices of $G$.
	If some prescribed vertices $v_i$ of $G$ are given and the goal is to find a solution to the multiway partition problem so that the cluster $V_i$ contains the vertex $v_i$, the problem becomes harder. It is proved to be {\bf NP-hard} for $r=3$ and the maximum degree in $G$ is at most $4$. On the positive side, this restriction can be solved in polynomial time when the input graph is planar. However, if $r$ is arbitrary then the problem is {\bf NP-hard} even when $G$ is planar \cite{Dahlhaus94thecomplexity}.
	
	The case of the multiway partition problem, in which $r=2$, is frequently encountered in literature. This case is called the bipartition problem. It is {\bf NP-hard} for $d$-regular graphs \cite{BuiChauduri1}, where $d\geq 3$ is a fixed constant. On the positive side, there is a dynamical programming based algorithm for solving this problem in the class of trees \cite{Bui2,Goldberg1,MacGregor1}. Note that the computational complexity of the bipartition problem is unknown if $G$ is a planar graph.

	\section{Computational Complexity of {\sc CN}}
	\label{sec:comp}
		
	In this section, we obtain the main results that deal with the computational complexity of {\sc CN}. We prove theorems that establish the {\bf NP-hardness} of some variants of {\sc CN}. Our reductions imply that {\sc CN} is inapproximable within a certain factor.
	
	In order to formulate the results, we consider {\sc CNWD}, which is formulated as follows:\\
	
	{\sc CNWD:} Given a DAG $G=(V,E)$, with vertex-weight function $w: V \rightarrow \mathbb{N}$, delay function $\delta: V \rightarrow \mathbb{N}$, constants $d$ and $D$,  cluster capacity $M$ and a positive integer $k$, partition $V$ into clusters such that
	\begin{enumerate}
		\item  The weight of each cluster is bounded by $M$,
		\item The length of an optimal path of $G$ is at most $k$.
	\end{enumerate}
	
	\begin{wrapfigure}[20]{R}{5cm}
		\begin{tikzpicture}[->,>=stealth',shorten >=1pt,auto,node distance=3cm,
		thick,main node/.style={circle,fill=blue!20,draw,font=\sffamily\LARGE\bfseries}]
		
		\node[circle,fill=blue!20,draw] at (0,0) (s) {$s$};
		
		\node[circle,fill=blue!20,draw,scale=0.7] at (-2,-2) (v1) {$v_1$};
		
		\node[circle,fill=blue!20,draw,scale=0.7] at (-1,-2) (v2) {$v_2$};
		
		\node[circle,fill=blue!20,draw,scale=0.7] at (2,-2) (vn) {$v_n$};
		
		\node[circle,fill=blue!20,draw] at (0,-4) (t) {$t$};

		\path[every node]
		(s) edge  (v1)
		edge  (v2)
		edge  (vn)
		
		(v1) edge (t)
		(v2) edge (t)
		(vn) edge (t)
		
		(v2) -- node[auto=false]{\ldots} (vn);
		
		\end{tikzpicture}
		
		
		\caption{Reduction from the {\sc Partition} problem to {\sc CNWD}$\langle W, M, \Delta \rangle$.}\label{npcompleteness}
	\end{wrapfigure}
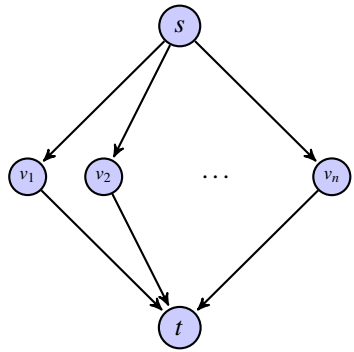
	
	It is not hard to see that {\sc CNWD} is the decision version of {\sc CN}$\langle W, M, \Delta \rangle$. We make this correspondence explicit by writing {\sc CNWD} as {\sc CNWD}$\langle W, M, \Delta \rangle$. We use the same notation for restrictions of {\sc CN}$\langle W, M, \Delta \rangle$.
	
	Note that {\sc CNWD}$\langle W, M, \Delta \rangle$ is in {\bf NP}. This follows from the well-known fact that a maximum weighted path in an edge-weighted DAG can be found in polynomial time.
	
	If $A $ is a subset of positive integers, then we denote by {\sc CNWD}$\langle A, M, \Delta \rangle$, the restriction of {\sc CNWD}$\langle W, M, \Delta \rangle$, when the weights of verticies of the input DAG are from $A$.
	
	Our first theorem establishes the {\bf NP-completeness} of \mbox{{\sc CNWD}$\langle W, M, \Delta \rangle$}. Clearly, this means that {\sc CN}$\langle W, M, \Delta \rangle$ is {\bf NP-hard}.

	\begin{theorem}\label{MainCompTheorem} {\sc CNWD}$\langle W, M, \Delta \rangle$ is {\bf NP-complete}.
	\end{theorem}

	\begin{proof} It is clear that {\sc CNWD}$\langle W, M, \Delta \rangle$ is in {\bf NP}. This follows from the well-known fact that a maximum weighted path in an edge-weighted DAG can be found in polynomial time.
		
		In order to establish {\bf NP-hardness} of {\sc CNWD}$\langle W, M, \Delta \rangle$, we present a reduction from the {\sc Partition} problem.
		
	Recall the {\sc Partition} problem:\\
		
		{\sc Partition:} Given a set $S=\{ a_{1}, a_{2}, \ldots, a_{n}\}$, the goal is to check whether there is a set $S_{1} \subset S$, such that $\sum_{x \in S_{1}} x  = \sum_{x \in S-S_{1}} x$. \\

		Without loss of generality, we assume that \mbox{$B=\sum_{i\in S}a_{i} $} is even, otherwise the problem is trivial.
		
		We now construct an instance $I'$ of {\sc CNWD}$\langle W, M, \Delta \rangle$ as shown in Figure \ref{npcompleteness}. There is a source $s$ connected to a sink $t$ through $n$ vertices labeled $v_{1}$ through $v_{n}$.
		None of the $v_{i}$ vertices are connected to each other. Each of the $v_{i}$ vertices has a weight $a_{i}$, and $s$ and $t$ have weights equal to $\frac{B}{2}$. We set $D=1$ and $d=0$. All vertices are given a delay of $0$. The cluster capacity is set to $B$, and we take $k=1$. The description of $I'$ is complete.
		
		We observe that $I'$ can be constructed from an instance $I$ of the {\sc Partition} problem in polynomial time. In order to complete the proof of the theorem, we show that $I$ is a ``yes" instance of the {\sc Partition} problem if and only if $I'$ is a ``yes" instance of {\sc CNWD}$\langle W, M, \Delta \rangle$.
		
		Assume that $I$ is a ``yes" instance of the {\sc Partition} problem. This means that there exists a partition of $S$ into $S_{1}$ and $S-S_{1}$ such that $\sum_{x \in S_{1}} x  = \sum_{x \in S-S_{1}} x = \frac{B}{2}$. Group the vertices corresponding to the elements in $S_{1}$ with $s$, and the remaining vertices with $t$ in $G$. Observe that the packing constraint is met. Moreover, the length of the optimal path from $s$ to $t$ is $1$. This means that $I'$ is a ``yes" instance of {\sc CNWD}$\langle W, M, \Delta \rangle$.
		
		For the proof of the converse statement, assume that $I'$ is a ``yes" instance of {\sc CNWD}$\langle W, M, \Delta \rangle$. This means that there is a way of packing the vertices of $G$ into clusters such that the length of the optimal path from $s$ to $t$ is $1$. We observe that every vertex must be packed with either $s$ or $t$, otherwise the length of the optimal path must equal $2$ going through that vertex. Let $w(s)$ and $w(t)$ be the weights of vertices $s$ and $t$, respectively, and let
		$w(s_{i})$ and $w(t_{i})$ be the sum of the weights of vertices packed with $s$ and $t$, respectively. Clearly,
		\begin{equation*}
			w(s)+w(s_{i})+w(t) + w(t_{i}) = 2\cdot B.
		\end{equation*} Since
		
		\begin{equation*}
			w(s)+w(s_{i})\le B \text{ and } w(t)+w(t_{i}) \le B,
		\end{equation*} we have
		
		\begin{equation*}
			w(s_{i}) \le \frac{B}{2} \text{ and } w(t_{i}) \le \frac{B}{2}.
		\end{equation*} This implies that
		
		\begin{equation*}
			w(s_{i}) = \frac{B}{2} \text{ and } w(t_{i}) = \frac{B}{2}.
		\end{equation*}Thus, we have obtained the desired partition of $S$. Hence, $I$ is a ``yes" instance of the {\sc Partition} problem.
		
		The proof of the theorem follows. $\blacksquare$
	\end{proof}
	
	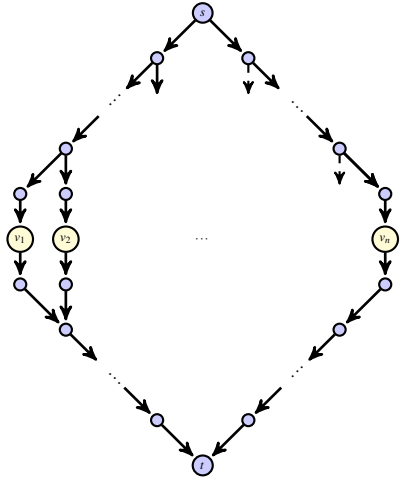
\begin{figure}[H]
		\begin{center}
		\begin{tikzpicture}[scale=0.6,->,>=stealth',shorten >=1pt,auto,node distance=1cm, thick,
		yellow node/.style={circle,fill=yellow!20,draw,font=\sffamily\small\bfseries, scale=.6},
		blue node/.style={circle,fill=blue!20,draw,font=\sffamily\small\bfseries, scale=.6},
		blank node/.style={circle,fill=none,draw=none,font=\sffamily\small\bfseries, scale=.6}]
		
		\node[blue node] (s) at (0,0) {$s$};
		\node[blank node] (blank1) [below left of = s] {};
		\node[blank node] (blank2) [below right of = s] {};
		\node[blue node] (s11) at (-1,-1) {};
		\node[blue node] (s12) at (1,-1) {};
		\node[blank node, rotate=135] (s21) at (-2,-2) {$\vdots$};
		\node[blank node] (s21a) at (-1,-2) {};
		\node[blank node] (s22) at (1,-2) {};
		\node[blank node, rotate=45] (s22a) at (2,-2) {$\vdots$};
		\node[blue node] (s31) at (-3,-3) {};
		\node[blue node] (s32) at (3,-3) {};
		
		\node[blue node] (s1) at (-4,-4) {};
		\node[blue node] (s2) at (-3,-4) {};
		\node[blank node] (sm) at (3,-4) {};
		\node[blue node] (sn) at (4,-4) {};
		\node[yellow node] (v1) at (-4,-5) {$v_1$};
		\node[yellow node] (v2) at (-3,-5) {$v_2$};
		\node[blank node] at (0,-5) {$\cdots$};
		\node[yellow node] (vn) at (4,-5) {$v_n$};
		\node[blue node] (t1) at (-4,-6) {};
		\node[blue node] (t2) at (-3,-6) {};
		\node[blue node] (tn) at (4,-6) {};
		
		\node[blue node] (t31) at (-3,-7) {};
		\node[blue node] (t32) at (3,-7) {};
		\node[blank node, rotate=45] (t21) at (-2,-8) {$\vdots$};
		\node[blank node, rotate=135] (t22) at (2,-8) {$\vdots$};
		\node[blue node] (t11) at (-1,-9) {};
		\node[blue node] (t12) at (1,-9) {};
		\node[blue node] (t) at (0,-10) {$t$};
		
		\foreach \source/ \dest in {s/s11, s/s12, s11/s21, s12/s22a, s11/s21a, s21/s31, s22a/s32, s31/s1, s31/s2, s32/sn, s32/sn, s1/v1, v1/t1, s2/v2, v2/t2, sn/vn, vn/tn, t1/t31, t2/t31, tn/t32, t31/t21, t32/t22, t21/t11, t22/t12, t11/t, t12/t}
		\path[edge] (\source) -- (\dest);
		
		\draw[dashed,->] (s32) -- (sm);
		\draw[dashed,->] (s12) -- (s22);
		
		\end{tikzpicture}
		
			\end{center}
		
		\caption{Reduction from the {\sc Partition} problem to {\sc CNWD}$\langle W, M, 3 \rangle$.} \label{CN_WM3_partitionreduction}
	\end{figure}
	
	The next theorem serves to strengthen Theorem \ref{MainCompTheorem}.
	
	\begin{theorem}\label{CN_WM3Theorem} {\sc CNWD}$\langle W, M, 3 \rangle$ is {\bf NP-complete}.
	\end{theorem}

	\begin{proof} It is clear that {\sc CNWD}$\langle W, M, 3 \rangle$ is in {\bf NP}. This follows from the well-known fact that a maximum weighted path in an edge-weighted DAG can be found in polynomial time.
		
		In order to establish {\bf NP-hardness} of {\sc CNWD}$\langle W, M, 3 \rangle$, we present a reduction from {\sc Partition}.
		
		We construct a new instance $I'$ of {\sc CNWD}$\langle W, M, 3 \rangle$ as shown in Figure \ref{CN_WM3_partitionreduction}.
		
		Each vertex $v_{i}$ ($i \in \{1, \ldots, n\}$) belongs to a path which connects the source $s$ to the sink $t$. Let $V$ denote the set of all $v_i$ vertices. Let $S$ denote the set of all vertices that are predecessors to the vertices in $V$, and let $T$ denote the set of all vertices that are successors to the vertices in $V$. Since $|S|=|T|$, let $m$ denote the size of $S$ and $T$. No pair of vertices in $V$ are connected. Each vertex $v_{i} \in V$ has a weight of $a_{i}$. Every vertex in $S$ and $T$ has weight $1$. So, the sum of the weights of all vertices in $S$ is equal to $m$, and the sum of the weights of all vertices in $T$ is equal to $m$. We set $D=1$ and $d=0$. Every vertex is given a delay of $0$. The cluster capacity $M$ is set to $\left( \frac{B}{2}+m \right)$, and we take $k=1$. The description of $I'$ is complete.
		
		Observe that $I'$ can be constructed from an instance $I$ of the {\sc Partition} problem in polynomial time. In order to complete the proof of the theorem, we show that $I$ is a ``yes" instance of {\sc Partition}, if and only if $I'$ is a ``yes" instance of {\sc CNWD}$\langle W, M, 3 \rangle$.
		
		Assume that $I$ is a ``yes" instance of {\sc Partition}. This means that there exists a partition of $A$ into $A_{1}$ and $A_{2}$, such that $\sum_{x \in A_{1} } x= \sum_{x \in A_{2} } x = \frac{B}{2}$. Group the vertices corresponding to the elements in $A_{1}$ with $S$, and the remaining vertices with $T$. Observe that the cluster capacity constraint is met. Moreover, the length of the optimal path from a source to a sink is $1$. This means that $I'$ is a ``yes" instance of {\sc CNWD}$\langle W, M, 3 \rangle$.
		
		Conversely, assume that $I'$ is a ``yes" instance of {\sc CNWD}$\langle W, M, 3 \rangle$. This means that there is a way of packing the vertices of the DAG in Figure \ref{CN_WM3_partitionreduction} into clusters, such that the cluster capacity is not exceeded, and the length of the optimal path from $s$ to $t$ is $1$.
		
		Observe that every vertex belonging to $S$ must be clustered together, and every vertex belonging to $T$ must be clustered together. Otherwise, the length of the path from $s$ to $t$ is greater than $1$. Additionally, if there is a vertex $v_i \in V$ which is not packed with either $S$ or $T$, then the length of the path from $s$ to $t$ is greater than $1$. Therefore, $V$ cannot be partitioned into more than two sets.
		
		Let $V_S$ and $V_T$ denote the subset of vertices $v_i \in V$ that are packed with $S$ and $T$, respectively. Observe that $V_S \cup V_T = V$. Moreover, the length of the path from $s$ to any vertex in $V_S$ must be $0$, and the length of the path from any vertex in $V_T$ to $t$ must also be $0$.

		Let $w(S)$ denote the sum of the weights of all vertices in $S$, and $w(T)$ denote the sum of the weights of all vertices in $T$. Notice that $w(S)=w(T)=m$. Let $w(V_{S})$ and $w(V_{T})$ denote the sum of the weights of all vertices in $V_S$ and $V_T$, respectively.
		
		Notice that,
		\begin{equation*}
			w(S)+w(V_{S})+w(T) + w(V_{T}) = B + 2 \cdot m.
		\end{equation*}
		
		Since \[ w(S)+w(V_{S})\le \left( \frac{B}{2}+m \right) \text{ and }\]
		\[ w(T)+w(V_{T}) \le \left( \frac{B}{2}+m \right), \]
		
		then \[w(V_{S}) \le \frac{B}{2} \text{ and }  w(V_{T}) \le \frac{B}{2}. \]
		
		This implies that
		
		\[w(V_{S}) = \frac{B}{2} \text{  and  } w(V_{T}) = \frac{B}{2}.\]
		
		Thus, we have obtained the desired partition of $A$. Hence, $I$ is a ``yes" instance of {\sc Partition}.
		
		The proof of the theorem follows. $\blacksquare$
	\end{proof}

	The proof of Theorem \ref{CN_WM3Theorem} implies an inapproximability result for {\sc CN}$\langle W, M, 3 \rangle$.
	
	\begin{corollary} {\sc CN}$\langle W, M, 3 \rangle$ does not admit a $(2-\varepsilon)$-approximation algorithm for each $\varepsilon >0$, unless {\bf P=NP}.
	\end{corollary}
	
	\begin{proof} Consider the reduction from {\sc Partition} described in the proof of theorem \ref{CN_WM3Theorem}. Observe that in any approximate solution of the clustering problem, there must exist at least one vertex which is not packed with $s$ or $t$. 
    This means there are at least $2$ $D$-edges along any source to sink path. Hence, any $(2-\varepsilon)$-approximation algorithm can be used to solve the partition problem exactly.
		
		The proof of the corollary follows. $\blacksquare$
	\end{proof}
	
	In the proof of the following theorem, we use a {\sc 3SAT} reduction modeled after the one presented in \cite{Kag03}.
	
	\begin{theorem}\label{CN_W123_M3_Delta3Theorem} {\sc CNWD}$\langle \{1,2,3\}, 3, 3 \rangle$ is {\bf NP-complete}.
	\end{theorem}
	
	\begin{proof} It is clear that {\sc CNWD}$\langle \{1,2,3\}, 3, 3 \rangle$ is in {\bf NP}. This follows from the well-known fact that a maximum weighted path in an edge-weighted DAG can be found in polynomial time.
		
		In order to establish {\bf NP-hardness} of {\sc CNWD}$\langle \{1,2,3\}, 3, 3 \rangle$, we reduced from {\sc 3SAT}.
		
		For that purpose, we recall {\sc 3SAT} as follows:
		
		{\sc 3SAT:} Given a $3$-CNF formula $\phi$ with $n$ variables $x_1, \ldots, x_n$ and $m$ clauses $C_1, \ldots, C_m$, the goal is to check whether $\phi$ has a satisfying assignment.
		
		Without loss of generality, for all $i \in \{1, \ldots, n\}$ we assume that each variable $x_i$ in $\phi$ appears at most $3$ times and each literal at most twice. (Any {\sc 3SAT} instance can be transformed to satisfy these properties in polynomial time \cite{Papa94}.)
		
		Let each variable $x_i$ $(1 \leq i \leq n)$, be represented by a variable gadget as shown in Figure \ref{gadgets} (a). Let each clause $C_j$ $(1 \leq j \leq m)$, be represented by a clause gadget as shown in Figure \ref{gadgets} (b). If a variable $x_i$ or its complement $\bar{x}_i$ is the $1$st, $2$nd, or $3$rd literal of a clause $C_j$, then the corresponding vertex labeled $x_i$ (or $\bar{x}_i$) is connected to a sink labeled $C_j$ through a pair of vertices labeled $y_{j1}$ and $z_{j1}$, $y_{j2}$ and $z_{j2}$, or $y_{j3}$ and $z_{j2}$, respectively.
		
		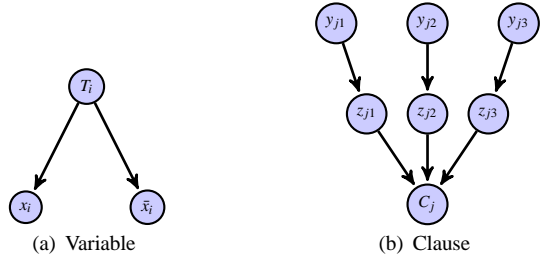
\begin{figure}[ht]
			\centering
			\subfigure[Variable]{
				\begin{tikzpicture}[scale=.8,->,>=stealth',shorten >=1pt,auto,node distance=1.25cm, thick,
				blue node/.style={circle,fill=blue!20,draw,font=\sffamily\small\bfseries, scale=.8},
				red node/.style={circle,fill=red!20,draw,font=\sffamily\Large\bfseries, scale=.8}]
				
				\node[blue node] (Ti) at (0,0) {$T_i$};
				\node[blue node] (xi) at (-1,-2) {$x_i$};
				\node[blue node] (notxi) at (1,-2) {$\bar{x}_i$};
				
				\foreach \source/ \dest in {Ti/xi, Ti/notxi}
				\path[edge] (\source) -- (\dest);
				
				\end{tikzpicture}}
			\hspace{50pt}
			\subfigure[Clause]{
				\begin{tikzpicture}[scale=.8,->,>=stealth',shorten >=1pt,auto,node distance=1.25cm, thick,
				blue node/.style={circle,fill=blue!20,draw,font=\sffamily\small\bfseries, scale=.8},
				red node/.style={circle,fill=red!20,draw,font=\sffamily\Large\bfseries, scale=.8}]
				
				\node[blue node] (Cj) at (0,0) {$C_j$};
				\node[blue node] (yj1) at (-1.5,3) {$y_{j1}$};
				\node[blue node] (yj2) at (0,3) {$y_{j2}$};
				\node[blue node] (yj3) at (1.5,3) {$y_{j3}$};
				\node[blue node] (zj1) at (-1,1.5) {$z_{j1}$};
				\node[blue node] (zj2) at (0,1.5) {$z_{j2}$};
				\node[blue node] (zj3) at (1,1.5) {$z_{j3}$};

				\foreach \source/ \dest in {yj1/zj1, yj2/zj2, yj3/zj3, zj1/Cj, zj2/Cj, zj3/Cj}
				\path[edge] (\source) -- (\dest);
				
				\end{tikzpicture}}
			\caption{Gadgets used to represent variables and clauses.} \label{gadgets}
		\end{figure}

		We now construct an instance $I'$ of {\sc CNWD}$\langle \{1,2,3\}, 3, 3 \rangle$ as shown in Figure \ref{CN_W123_M3_Delta3_3SATreduction1}.
		
		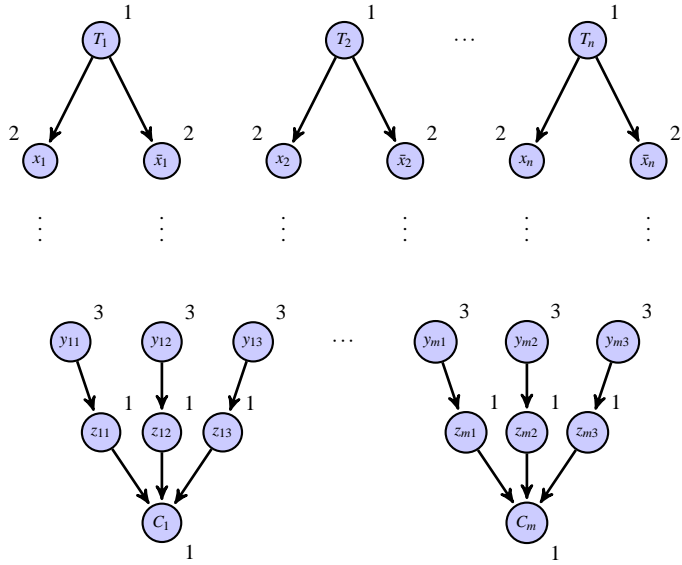
\begin{figure}[ht]
			\begin{center}
				\begin{tikzpicture}[scale=.8,->,>=stealth',shorten >=1pt,auto,node distance=1.25cm, thick,
				blue node/.style={circle,fill=blue!20,draw,font=\sffamily\small\bfseries, scale=.8},
				red node/.style={circle,fill=red!20,draw,font=\sffamily\Large\bfseries, scale=.8}]
				
				\node[blue node] (T1) at (-4,0) [label=above right:$1$] {$T_1$};
				\node[blue node] (x1) at (-5,-2) [label=above left:$2$] {$x_1$};
				\node[blue node] (notx1) at (-3,-2) [label=above right:$2$] {$\bar{x}_1$};
				\node[blue node] (T2) at (0,0) [label=above right:$1$] {$T_2$};
				\node[blue node] (x2) at (-1,-2) [label=above left:$2$] {$x_2$};
				\node[blue node] (notx2) at (1,-2) [label=above right:$2$] {$\bar{x}_2$};
				\node[draw=none] (ellipsis) at (2,0) {$\cdots$};
				\node[blue node] (Tn) at (4,0) [label=above right:$1$] {$T_n$};
				\node[blue node] (xn) at (3,-2) [label=above left:$2$] {$x_n$};
				\node[blue node] (notxn) at (5,-2) [label=above right:$2$] {$\bar{x}_n$};
				\node[draw=none] (ellipsis) at (-5,-3) {$\vdots$};
				\node[draw=none] (ellipsis) at (-3,-3) {$\vdots$};
				\node[draw=none] (ellipsis) at (-1,-3) {$\vdots$};
				\node[draw=none] (ellipsis) at (1,-3) {$\vdots$};
				\node[draw=none] (ellipsis) at (3,-3) {$\vdots$};
				\node[draw=none] (ellipsis) at (5,-3) {$\vdots$};
				\node[blue node] (C1) at (-3,-8) [label=below right:$1$] {$C_1$};
				\node[blue node] (y11) at (-4.5,-5) [label=above right:$3$] {$y_{11}$};
				\node[blue node] (y12) at (-3,-5) [label=above right:$3$] {$y_{12}$};
				\node[blue node] (y13) at (-1.5,-5) [label=above right:$3$] {$y_{13}$};
				\node[blue node] (z11) at (-4,-6.5) [label=above right:$1$] {$z_{11}$};
				\node[blue node] (z12) at (-3,-6.5) [label=above right:$1$] {$z_{12}$};
				\node[blue node] (z13) at (-2,-6.5) [label=above right:$1$] {$z_{13}$};
				\node[draw=none] (ellipsis) at (0,-5) {$\cdots$};
				\node[blue node] (Cm) at (3,-8) [label=below right:$1$] {$C_m$};
				\node[blue node] (ym1) at (1.5,-5) [label=above right:$3$] {$y_{m1}$};
				\node[blue node] (ym2) at (3,-5) [label=above right:$3$] {$y_{m2}$};
				\node[blue node] (ym3) at (4.5,-5) [label=above right:$3$] {$y_{m3}$};
				\node[blue node] (zm1) at (2,-6.5) [label=above right:$1$] {$z_{m1}$};
				\node[blue node] (zm2) at (3,-6.5) [label=above right:$1$] {$z_{m2}$};
				\node[blue node] (zm3) at (4,-6.5) [label=above right:$1$] {$z_{m3}$};
				
				\foreach \source/ \dest in {T1/x1, T1/notx1, T2/x2, T2/notx2, Tn/xn, Tn/notxn, y11/z11, y12/z12, y13/z13, z11/C1, z12/C1, z13/C1, ym1/zm1, ym2/zm2, ym3/zm3, zm1/Cm, zm2/Cm, zm3/Cm}
				\path[edge] (\source) -- (\dest);
				
				\end{tikzpicture}
				
			\end{center}
			
			\caption{Reduction from the {\sc 3SAT} problem to {\sc CNWD}$\langle \{1,2,3\}, 3, 3 \rangle$.} \label{CN_W123_M3_Delta3_3SATreduction1}
		\end{figure}

		The resulting DAG $G$ represents a combinatorial circuit. Let $V$ denote the set of all vertices labeled $x_i$ or $\bar{x}_i$ $(1 \leq i \leq n)$. There are $n$ sources $T_i$ $(1 \leq i \leq n)$ connected to $m$ sinks $C_j$ $(1 \leq j \leq m)$ through some vertices in $V$ and $3m$ pairs of vertices labeled $y_{jp}$ and $z_{jp}$ $(1 \leq j \leq m, 1 \leq p \leq 3)$. Each $y_{jp}$ is connected to exactly one vertex gadget, and for fixed $j$, no two vertices in $\{y_{j1}, y_{j2}, y_{j3}\}$ are adjacent to the vertices $x_i$ and $\bar{x}_i$ belonging to the same vertex gadget. In other words,  $x_i$ and $\bar{x}_i$ cannot both be connected to the same clause gadget. Every $T_{i}$, $z_{jp}$, and $C_j$ has a weight of $1$, every $x_i, \bar{x}_i \in V$ has a weight of $2$, and every $y_{jp}$ has a weight of $3$. We set $D=1$ and $d=0$. All vertices are given a delay of $0$. The cluster capacity $M$ is set to $3$, and we take $k=3$. The description of $I'$ is complete.
		
		Observe that $I'$ can be constructed from $I$ in polynomial time. In order to complete the proof of the theorem, we show that $I$ is a ``yes" instance of {\sc 3SAT}, if and only if $I'$ is a ``yes" instance of {\sc CNWD}$\langle \{1,2,3\}, 3, 3 \rangle$.
		
		Suppose that $I$ is a ``yes" instance of {\sc 3SAT}. This means that there exists an assignment of $\phi$ such that every clause has at least one true literal. If a literal is set to {\bf true}, then the corresponding vertex $x_i$  (or $\bar{x}_i$) should be clustered with $T_i$, but if it is set to {\bf false}, then the corresponding vertex is clustered alone. Notice that every $y_{jp}$ must be clustered alone. Since each clause $C_j$ has at least one true literal, the vertex $z_{jp}$ corresponding to that literal should be clustered alone. This means that the source to sink path going through vertices $y_{jp}$ and $z_{jp}$ corresponding to {\bf true} literals have length $3$. If either of the other two $z_{jp}$ vertices belonging to the respective clause gadget corresponds to literals which are set to {\bf false}, they should be clustered with $C_j$. Otherwise, they may also be clustered alone. Note that clustering two $z_{jp}$ vertices with $C_j$, even if they both correspond to {\bf true} literals, leads to paths of length $2 < 3=k$. Observe that the cluster capacity constraint is met, and the length of the optimal path from any source $T_i$ to any sink $C_j$ is $3$. This means that $I'$ is a ``yes" instance of {\sc CNWD}$\langle \{1,2,3\}, 3, 3 \rangle$.
		
		Conversely, suppose that $I'$ is a ``yes" instance of {\sc CNWD}$\langle \{1,2,3\}, 3, 3 \rangle$. This means that there is a way of packing the vertices of $G$ into clusters of capacity $M=3$, such that the length of the optimal path from source to sink is $3$.
		
		Since $M=3$, again notice that every $y_{jp}$ must be clustered alone. Each vertex $C_j$ may be clustered with at most $2$ of the $z_{jp}$ vertices. So, at least one $z_{jp}$ is clustered alone. However, notice that any source to sink path with a vertex $z_{jp}$ clustered alone, has length at least $3$. In order to satisfy the optimal path constraint, each $T_i$ must be clustered with the vertex $x_i$ (or $\bar{x}_i$) which corresponds to a vertex $z_{jp}$ clustered alone. Otherwise, the length of the path would be $4 > k=3$. To avoid exceeding the cluster capacity, either $x_i$ or $\bar{x}_i$ (but not both) may be clustered with $T_i$. Finally, notice that $z_{jp}$ vertices along paths where their corresponding literals are considered {\bf false}, must be clustered with their respective sinks $C_j$. Otherwise, the length of such a path with all internal vertices belonging to single vertex clusters would have length $4 > 3=k$. Take the variable which corresponds to the vertex clustered with $T_i$, and set its value to {\bf true}. Take the variable which corresponds to the vertex not clustered with $T_i$, and set its value to {\bf false}.
		
		By setting to {\bf true} all literals with corresponding vertices $x_i$ (or $\bar{x}_i$) clustered with $T_i$, and by setting to {\bf false} all literals with corresponding vertices not clustered with $T_i$, means that at least one {\bf true} literal appears in every clause. Thus, a satisfying clustering for $G$ yields a satisfying assignment for $\phi$. Hence, $I$ is a ``yes" instance of {\sc 3SAT}.
		
		The proof of the theorem follows. $\blacksquare$
	\end{proof}
	
	The proof of Theorem \ref{CN_W123_M3_Delta3Theorem} implies an inapproximability result for {\sc CN}$\langle \{1,2,3\}, 3, 3 \rangle$.
	
	\begin{corollary} {\sc CN}$\langle \{1,2,3\}, 3, 3 \rangle$ does not admit a $(\frac{4}{3}-\varepsilon)$-approximation algorithm for each $\varepsilon >0$, unless {\bf P=NP}.
	\end{corollary}
	
	\begin{proof} Consider the reduction from {\sc 3SAT} described in the proof of Theorem \ref{CN_W123_M3_Delta3Theorem}. Observe that in any approximate solution of the clustering problem, there are at least $3$ and at most $4$ $D$-edges along any source to sink path. \comment{This means that $s$ and $t$ do not belong to the same cluster.} Hence, any $(\frac{4}{3} - \varepsilon)$-approximation algorithm can be used to solve the {\sc 3SAT} problem exactly.
		
		The proof of the corollary follows. $\blacksquare$
	\end{proof}
	
	\comment{We were able to obtain one more {\bf NP-hardness} result. More precisely, we were able to show that {\sc CN}$\langle \{1,2\}, 2, 4 \rangle$ is {\bf NP-hard}.}
	
	The next theorem is a restriction of {\sc CNWD}$\langle W, 2, \Delta \rangle$.
	
	\begin{theorem}\label{CN_W12_M2_Delta4Theorem20150713} {\sc CNWD}$\langle \{1,2\}, 2, 4 \rangle$ is {\bf NP-complete}.
	\end{theorem}

	\begin{proof}  It is clear that {\sc CNWD}$\langle \{1,2\}, 2, 4 \rangle$ is in {\bf NP}. This follows from the well-known fact that a maximum weighted path in an edge-weighted DAG can be found in polynomial time.
		
		In order to establish {\bf NP-hardness} of {\sc CNWD}$\langle \{1,2\}, 2, 4 \rangle$, we present a reduction from {\sc 3-Bounded Positive 1-in-3 SAT} ({\sc 3-BP 1-in-3 SAT}).
		
		\comment{Our proof uses a reduction from the following problem.\\}
		
		For that purpose, we recall {\sc 3-BP 1-in-3 SAT} as follows:
		
		{\sc 3-BP 1-in-3 SAT}: We are given a $3$-CNF formula $\phi$ with $n$ positive variables $x_1, \ldots, x_n$ and $m$ clauses $C_1, \ldots, C_m$, such that each variable appears in at most $3$ clauses. The goal is to check whether $\phi$ has a satisfying assignment such that every clause of $\phi$ has exactly one {\bf true} literal  \cite{Den09}.
		
		Let each variable $x_i$ $(1 \leq i \leq n)$, be represented by a variable gadget as shown in Figure \ref{gadgets2} (a). Let each clause $C_j$ $(1 \leq j \leq m)$, be represented by a clause gadget as shown in Figure \ref{gadgets2} (b). If a variable $x_i$ is the $1$st, $2$nd, or $3$rd literal of a clause $C_j$, then the corresponding vertex labeled $x_i$ is connected to a sink labeled $C_j$ through a pair of vertices labeled $y_{j1}$ and $z_{j1}$, $y_{j2}$ and $z_{j2}$, or $y_{j3}$ and $z_{j2}$, respectively.
		
		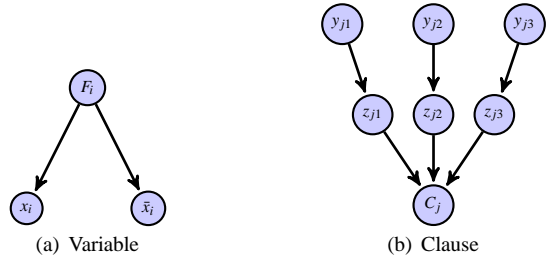
\begin{figure}[ht]
			\centering
			\subfigure[Variable]
			{
				\begin{tikzpicture}[scale=.8,->,>=stealth',shorten >=1pt,auto,node distance=1.25cm, thick,
				blue node/.style={circle,fill=blue!20,draw,font=\sffamily\small\bfseries, scale=.8},
				red node/.style={circle,fill=red!20,draw,font=\sffamily\Large\bfseries, scale=.8}]
				
				\node[blue node] (Ti) at (0,0) {$F_i$};
				\node[blue node] (xi) at (-1,-2) {$x_i$};
				\node[blue node] (notxi) at (1,-2) {$\bar{x}_i$};
				
				\foreach \source/ \dest in {Ti/xi, Ti/notxi}
				\path[edge] (\source) -- (\dest);
				
				\end{tikzpicture}
			}
			\hspace{50pt}
			\centering
			\subfigure[Clause]
			{		\begin{tikzpicture}[scale=.8,->,>=stealth',shorten >=1pt,auto,node distance=1.25cm, thick,
				blue node/.style={circle,fill=blue!20,draw,font=\sffamily\small\bfseries, scale=.8},
				red node/.style={circle,fill=red!20,draw,font=\sffamily\Large\bfseries, scale=.8}]
				
				\node[blue node] (Cj) at (0,0) {$C_j$};
				\node[blue node] (yj1) at (-1.5,3) {$y_{j1}$};
				\node[blue node] (yj2) at (0,3) {$y_{j2}$};
				\node[blue node] (yj3) at (1.5,3) {$y_{j3}$};
				\node[blue node] (zj1) at (-1,1.5) {$z_{j1}$};
				\node[blue node] (zj2) at (0,1.5) {$z_{j2}$};
				\node[blue node] (zj3) at (1,1.5) {$z_{j3}$};

				\foreach \source/ \dest in {yj1/zj1, yj2/zj2, yj3/zj3, zj1/Cj, zj2/Cj, zj3/Cj}
				\path[edge] (\source) -- (\dest);
				
				\end{tikzpicture}
			}
			\caption{Gadgets used to represent variables and clauses.} \label{gadgets2}
		\end{figure}

		We now construct an instance $I'$ of {\sc CNWD}$\langle \{1,2\}, 2, 4 \rangle$ as shown in Figure \ref{CN_W12_M2_Delta4_3BP1in3SATreduction}.
		
		\begin{figure}[ht]
			\begin{center}
				\begin{tikzpicture}[scale=.8,->,>=stealth',shorten >=1pt,auto,node distance=1.25cm, thick,
				blue node/.style={circle,fill=blue!20,draw,font=\sffamily\small\bfseries, scale=.8},
				red node/.style={circle,fill=red!20,draw,font=\sffamily\Large\bfseries, scale=.8}]
				
				\node[blue node] (T1) at (-4,0) [label=above right:$1$] {$F_1$};
				\node[blue node] (x1) at (-5,-2) [label=above left:$1$] {$x_1$};
				\node[blue node] (notx1) at (-3,-2) [label=above right:$1$] {$\bar{x}_1$};
				\node[blue node] (T2) at (0,0) [label=above right:$1$] {$F_2$};
				\node[blue node] (x2) at (-1,-2) [label=above left:$1$] {$x_2$};
				\node[blue node] (notx2) at (1,-2) [label=above right:$1$] {$\bar{x}_2$};
				\node[draw=none] (ellipsis) at (2,0) {$\cdots$};
				\node[blue node] (Tn) at (4,0) [label=above right:$1$] {$F_n$};
				\node[blue node] (xn) at (3,-2) [label=above left:$1$] {$x_n$};
				\node[blue node] (notxn) at (5,-2) [label=above right:$1$] {$\bar{x}_n$};
				\node[draw=none] (ellipsis) at (-5,-3) {$\vdots$};
				\node[draw=none] (ellipsis) at (-1,-3) {$\vdots$};
				\node[draw=none] (ellipsis) at (3,-3) {$\vdots$};
				\node[blue node] (C1) at (-3,-8) [label=below right:$1$] {$C_1$};
				\node[blue node] (y11) at (-4.5,-5) [label=above right:$2$] {$y_{11}$};
				\node[blue node] (y12) at (-3,-5) [label=above right:$2$] {$y_{12}$};
				\node[blue node] (y13) at (-1.5,-5) [label=above right:$2$] {$y_{13}$};
				\node[blue node] (z11) at (-4,-6.5) [label=above right:$1$] {$z_{11}$};
				\node[blue node] (z12) at (-3,-6.5) [label=above right:$1$] {$z_{12}$};
				\node[blue node] (z13) at (-2,-6.5) [label=above right:$1$] {$z_{13}$};
				\node[draw=none] (ellipsis) at (0,-5) {$\cdots$};
				\node[blue node] (Cm) at (3,-8) [label=below right:$1$] {$C_m$};
				\node[blue node] (ym1) at (1.5,-5) [label=above right:$2$] {$y_{m1}$};
				\node[blue node] (ym2) at (3,-5) [label=above right:$2$] {$y_{m2}$};
				\node[blue node] (ym3) at (4.5,-5) [label=above right:$2$] {$y_{m3}$};
				\node[blue node] (zm1) at (2,-6.5) [label=above right:$1$] {$z_{m1}$};
				\node[blue node] (zm2) at (3,-6.5) [label=above right:$1$] {$z_{m2}$};
				\node[blue node] (zm3) at (4,-6.5) [label=above right:$1$] {$z_{m3}$};
				
				\foreach \source/ \dest in {T1/x1, T1/notx1, T2/x2, T2/notx2, Tn/xn, Tn/notxn, y11/z11, y12/z12, y13/z13, z11/C1, z12/C1, z13/C1, ym1/zm1, ym2/zm2, ym3/zm3, zm1/Cm, zm2/Cm, zm3/Cm}
				\path[edge] (\source) -- (\dest);
				
				\end{tikzpicture}
				
			\end{center}
			
			\caption{Reduction from {\sc 3-BP 1-in-3 SAT} to {\sc CNWD}$\langle \{1,2\}, 2, 4 \rangle$.} \label{CN_W12_M2_Delta4_3BP1in3SATreduction}
		\end{figure}
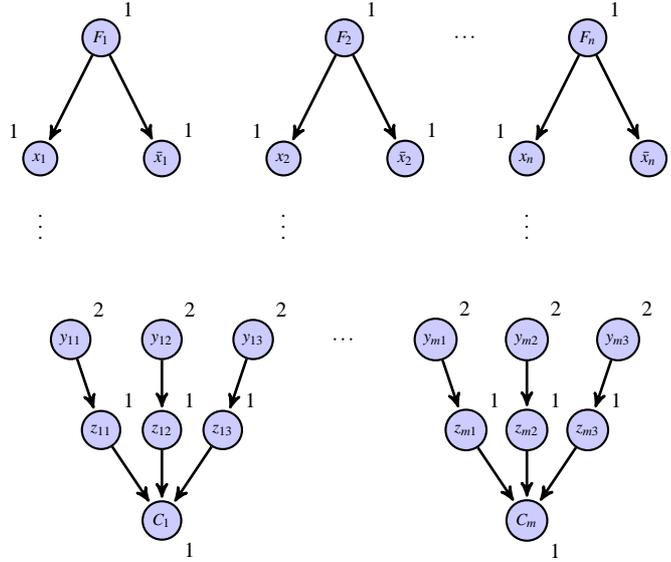

		The resulting DAG $G$ represents a combinatorial circuit. Let $V$ denote the set of all vertices labeled $x_i$ or $\bar{x}_i$ $(1 \leq i \leq n)$. There are $n$ sources $F_i$ $(1 \leq i \leq n)$ connected to $m$ sinks $C_j$ $(1 \leq j \leq m)$ through some vertices in $V$ and $3\cdot m$ pairs of vertices labeled $y_{jp}$ and $z_{jp}$ $(1 \leq j \leq m, 1 \leq p \leq 3)$. Each $y_{jp}$ is connected to exactly one vertex gadget. Every $x_i, \bar{x}_i \in V$, every $F_{i}$, $z_{jp}$, and $C_j$ has a weight of $1$. Every $y_{jp}$ has a weight of $2$. We set $D=1$ and $d=0$. All vertices are given a delay of $0$. The cluster capacity $M$ is set to $2$, and we take $k=3$. The description of $I'$ is complete.
		
		Observe that $I'$ can be constructed from $I$ in polynomial time. In order to complete the proof of the theorem, we show that $I$ is a ``yes" instance of {\sc 3-BP 1-in-3 SAT}, if and only if $I'$ is a ``yes" instance of {\sc CNWD}$\langle \{1,2\}, 2, 4 \rangle$.
		
		Suppose that $I$ is a ``yes" instance of {\sc 3-BP 1-in-3 SAT}. This means that there exists an assignment of $\phi$ such that every clause has exactly one {\bf true} literal. If a literal is set to {\bf true}, then the corresponding vertex $x_i$ should be clustered alone, but if it is set to {\bf false}, then the corresponding vertex is clustered with $F_i$. Since $M=2$, every $y_{jp}$ must be clustered alone. Since each clause $C_j$ has exactly one {\bf true} literal, the vertex $z_{jp}$ corresponding to that literal should be clustered with $C_j$. The other two $z_{jp}$ vertices belonging to the respective clause gadget should be clustered alone. Observe that the cluster capacity constraint is met, and the length of the optimal path from any source $F_i$ to any sink $C_j$ is $3$. This means that $I'$ is a ``yes" instance of {\sc CNWD}$\langle \{1,2\}, 2, 4 \rangle$.
		
		Conversely, suppose that $I'$ is a ``yes" instance of {\sc CNWD}$\langle \{1,2\}, 2, 4 \rangle$. This means that there is a way of packing the vertices of $G$ into clusters of capacity $M=2$ such that the length of the optimal path from source to sink is $3$.
		
		Since $M=2$, again notice that every $y_{jp}$ must be clustered alone. Each vertex $C_j$ may be clustered with at most $1$ of the $z_{jp}$ vertices. So, at least two of the $z_{jp}$ vertices are clustered alone. However, notice that any source to sink path with a vertex $z_{jp}$ clustered alone, has length at least $3$. In order to satisfy the optimal path constraint, each $F_i$ must be clustered with the vertex $x_i$ corresponding to a vertex $z_{jp}$ which is clustered alone. Otherwise, the length of the path would be $4 > 3=k$. So as not to exceed the cluster capacity, $F_i$ may be clustered with either $x_i$ or $\bar{x}_i$ (but not both). Finally, notice that all $z_{jp}$ vertices along paths with an isolated $x_i$ must be clustered with $C_j$. Otherwise, the length of such a path would have length $4 > 3=k$.
		
		By setting each literal whose corresponding vertex $x_i$ appears in the same cluster with $F_i$ to {\bf false}, and by setting each literal whose corresponding vertex $x_i$ is not clustered with $F_i$ to {\bf true}, we have that at least one {\bf true} literal in every clause. Thus, a satisfying clustering for $G$ yields a satisfying assignment for $\phi$. Hence, $I$ is a ``yes" instance of {\sc 3-BP 1-in-3 SAT}.
		
		The proof of the theorem follows. $\blacksquare$
	\end{proof}
	
	The proof of Theorem \ref{CN_W12_M2_Delta4Theorem20150713} implies an inapproximability result for {\sc CN}$\langle \{1,2\}, 2, 4 \rangle$.
	
	\begin{corollary} {\sc CN}$\langle \{1,2\}, 2, 4 \rangle$ does not admit a $(\frac{4}{3}-\varepsilon)$-approximation algorithm for each $\varepsilon >0$, unless {\bf P=NP}.
	\end{corollary}
	
	\begin{proof} Consider the reduction from {\sc 3-BP 1-in-3 SAT} described in the proof of Theorem \ref{CN_W12_M2_Delta4Theorem20150713}. Observe that in any approximate solution of the clustering problem, there are at least $3$ and at most $4$ $D$-edges along any source to sink path. Hence, the approximation algorithm can be used to solve the {\sc 3-BP 1-in-3 SAT} problem exactly.
		
		The proof of the corollary follows. $\blacksquare$
	\end{proof}

	\section{A $3$-Approximation Algorithm for {\sc CN}$\langle N, 2, \Delta \rangle$}
	\label{sec:ApproxResults}
	
	In this section, we present a $3$-approximation algorithm for {\sc CN}$\langle N, 2, \Delta \rangle$. Our algorithm makes use of the fact that there is a polynomial-time algorithm for finding a path with a maximum number of edges in DAGs. In each iteration, the algorithm picks a path $P$ with a maximum number of edges. Then it considers the central edge $e=(u,v)$ of $P$, and puts $u$ and $v$ in the same cluster. After that $u$ and $v$ are removed from $G$. The algorithm iterates until all edges of the input DAG are exhausted.
	
	\begin{theorem}\label{thm:approx} Algorithm \ref{algIter11257344} is a $3$-approximation algorithm for {\sc CN}$\langle N, 2, \Delta \rangle$.
	\end{theorem}
	
	\begin{algorithm}[htbp]
		
		\caption{A $3$-approximation algorithm for the clustering problem.}\label{algIter11257344}
		\begin{algorithmic}[1]
			\STATE Input: a DAG $G$;
			\STATE Output: a clustering of vertices of $G$;
			
			\STATE Take a longest path $P$ in the DAG $G$;
			\STATE Declare the central edge $e_{\lceil \frac{l}{2} \rceil}$ of $P$ as a $d$-edge, where $l$ denotes the length of $P$, and $e_1, \cdots , e_l$ are the edges of $P$.
			\STATE The edges adjacent to $e_{\lceil \frac{l}{2} \rceil}$ should be declared as $D$-edges.
			\STATE Remove edge $e_{\lceil \frac{l}{2} \rceil}$ together with its adjacent edges.
			\STATE Continue this process until all edges of $G$ are exhausted.
			
		\end{algorithmic}
	\end{algorithm}
	
	\begin{proof} For a path $P$, let $l(P)$ be the length of $P$ (i.e., the number of edges of $P$). Moreover, let
		\begin{equation*}
			l=\max_{P} l(P).
		\end{equation*} So, $l$ denotes the length of a longest path of $G$.
		
		The following shows a lower bound for $OPT$, where $OPT$ is the delay of the optimal clustering of $G$ when $M=2$.
		\begin{equation*}
			OPT \ge \lceil \frac{l(P)}{2} \rceil \cdot d+ \lfloor \frac{l(P)}{2} \rfloor \cdot D.
		\end{equation*} Since $P$ represents any path, then the above inequality must also be true for the longest path. Thus,
		\begin{equation*}
			OPT \ge \lceil \frac{l}{2} \rceil \cdot d+ \lfloor \frac{l}{2} \rfloor \cdot D.
		\end{equation*}
		
		Now, let us estimate $ALG$, where $ALG$ is the delay of the clustering found by the algorithm. We will consider $3$ cases. \vspace{8pt}
		
		Case 1: $l=1$. Then it can be easily seen that $ALG=OPT$.
		
		Case 2: $l$ is even. Then
		\begin{align*}
			ALG & \leq l\cdot D\\
			&\le 2\cdot (\lceil \frac{l}{2} \rceil \cdot d+ \lfloor \frac{l}{2} \rfloor \cdot D)\\
			&\le 2\cdot OPT\\
			&< 3\cdot OPT.
		\end{align*}
		
		Case 3: $l$ is odd and $l\geq 3$. Then
		\begin{align*}
			ALG &\leq l\cdot D\\
			&\le 3 \cdot \frac{l-1}{2} \cdot D\\
			&= 3 \cdot \lfloor \frac{l}{2} \rfloor \cdot D\\
			&\le 3\cdot (\lceil \frac{l}{2} \rceil \cdot d+ \lfloor \frac{l}{2} \rfloor \cdot D)\\
			&\le 3\cdot OPT.
		\end{align*}

		The proof of the theorem follows. $\blacksquare$
	\end{proof}
	
\begin{figure}[ht]
		\begin{center}
			\begin{tikzpicture}

			
			\node[circle,fill=black,draw] at (-2,2) (n2) {};
			
			\node[circle,fill=black,draw] at (0,2) (n3) {};
			
			\node[circle,fill=black,draw] at (0,3) (n33) {};
			\node[circle,fill=black,draw] at (0,4) (n333) {};
			
			\node[circle,fill=black,draw] at (2,2) (n4) {};
			\node[circle,fill=black,draw] at (2,1) (n44) {};
			\node[circle,fill=black,draw] at (2,0) (n444) {};
			
			\node[circle,fill=black,draw] at (4,2) (n5) {};

			\draw[->, thick] (n2)--(n3);
			
			\draw[->, thick] (n3)--(n4);
			\draw[->, thick] (n3)--(n33);
			\draw[->, thick] (n33)--(n333);
			
			\draw[->, thick] (n444)--(n44);
			\draw[->, thick] (n44)--(n4);
			\draw[->, thick] (n4)--(n5);
			
			\end{tikzpicture}

		\end{center}
		
		\caption{A DAG which obtains a factor $3$ approximation.}\label{2Boundcentfig11}
	\end{figure}
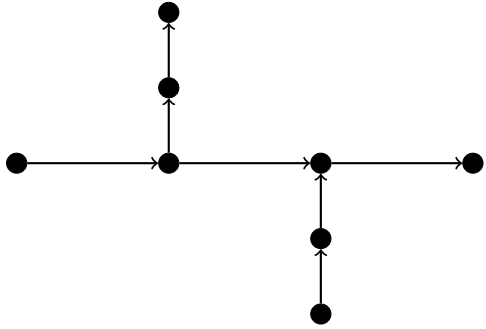

	Figure \ref{2Boundcentfig11} shows an example of a DAG for which the algorithm achieves an approximation factor of $3$.
	
	Observe that in this example, $OPT=2\cdot d+D$ and $ALG= 3\cdot D$. Hence, if $d=0$, we have $ALG=3\cdot OPT$.

	\section{Conclusion}
	\label{sec:conc}

	In this paper, we studied the problem of clustering combinatorial networks for delay minimization when logic replication is not allowed ({\sc CN}$\langle X, M, \Delta \rangle$). We showed that several versions of {\sc CN}$\langle W, M, \Delta \rangle$ are {\bf NP-hard}. The strategy developed for the proofs allowed us to prove that the problem does not admit a $(2-\varepsilon)$-approximation algorithm for any $\varepsilon >0$, unless {\bf P=NP}. On the positive side, there exists a $3$-approximation algorithm for {\sc CN}$\langle N, 2, \Delta \rangle$.
	
	\vspace{10pt}
	
	We are interested in the following open problems:
	\begin{enumerate}
		\item Finding an approximation algorithm for {\sc CN}$\langle N, 2, \Delta \rangle$ whose performance ratio is smaller than $3$. There may exist a combinatorial approximation algorithm for {\sc CN}$\langle N, 2, \Delta \rangle$ with smaller performance ratio. The following idea may be helpful in the design of such an algorithm. Take a longest path in the input DAG. Put the first two vertices in one cluster, the second two in another cluster, and so on. Remove all the vertices that are clustered with some other vertex. Iterate until all edges of the DAG are exhausted.
		

		\item Finding inapproximabilility results for several variants of {\sc CN}$\langle N, M, \Delta \rangle$. In particular, we are currently working on the decision problem {\sc CNWD}$\langle N, 2, 4 \rangle$ (a restriction of {\sc CNWD}$\langle N, M, \Delta \rangle$). It might be the case that {\sc CN}$\langle N, M, \Delta \rangle$ is {\bf NP-hard}. Moreover, it may not admit an FPTAS. 

		
	\end{enumerate}
	
	\medskip
	
	\paragraph{Acknowledgement:} The second author is indebted to his student Armen Davtyan, who has constructed the DAG from Figure \ref{2Boundcentfig11} showing that the performance ratio of the Algorithm \ref{algIter11257344} can be as large as $3$.

\end{document}